# Short-Term Environmental Effects and their Influence on Spatial Homogeneity of Organic Solar Cell Functionality

*Huei -Ting Chien[1*], Peter W. Zach[2], Bettina Friedel[1,3]*

[1]Institute of Solid State Physics, Graz University of Technology, Austria

[2]Institute of Analytical Chemistry and Food Chemistry, Graz University of Technology, Austria

[3]Energy Research Center, Vorarlberg University of Applied Sciences, Austria

*Email: hchien@tugraz.at

ABSTRACT

In this study, we focus on induced degradation and spatial inhomogeneity of organic photovoltaic devices under different environmental conditions, uncoupled from the influence of any auxiliary hole-transport (HT) layer. During testing of according devices comprising the standard photoactive layer of poly(3-hexylthiophene) (P3HT) as donor, blended with phenyl-$C_{61}$-butyric acid methyl ester ($PC_{60}BM$) as acceptor, a comparison was made between non-encapsulated devices upon exposure to argon in the dark, dry air in the dark, dry air with illumination and humid air in the dark. The impact on the active layer's photophysics is discussed, along with the device physics in terms of integral solar cell performance and spatially resolved photocurrent distribution with point-to-point analysis of the diode characteristics to determine the origin of the observed integrated OPV device behavior. The results show that even without widely used hygroscopic HT layer, poly(3,4-ethylenedioxythiophene):poly(styrenesulfonate) (PEDOT:PSS), humidity is still one major issue in short-term environmental degradation for organic solar cells with this architecture, and not only oxygen or light as often reported. Different from previous reports where water-induced device degradation was spatially homogeneous and a formation of $Al_2O_3$ islands only seen for oxygen permeation through pin holes in aluminum, we observed insulating islands merely after humidity exposure in the present study. Further, we demonstrated with laser beam induced current mapping and point-to-point diode analysis that the water-induced performance losses are a result of the exposed device area comprising regions with entirely unaltered high output and intact diode behavior and those with severe degradation showing detrimentally lowered output and voltage-independent charge blocking, practically insulating behavior. It is suggested that this is caused by transport of water through pin holes to the organic/metal interface where they form insulating oxide or hydroxide islands, while the organic active layer stays unharmed.

INTRODUCTION

Organic photovoltaic cells (OPVs) are considered a promising low-cost alternative source of renewable energy compared to traditional inorganic solar cells, due to their advantages of easy fabrication, low manufacturing cost, light weight, mechanical flexibility and environmental friendliness.[1-4] Due to the short diffusion range of photogenerated excitons in organic semiconductors, the most commonly used photoactive layer structure in OPVs is the donor/acceptor bulk-heterojunction (BHJ), basically consisting of a blend of donor and acceptor. One intensively studied BHJ system are blends of the conjugated polymer poly(3-hexylthiophene) (P3HT) for its excellent carrier mobility and a highly ordered packing structure[5,6] and the fullerene acceptor [6,6]-phenyl-$C_{61}$-butyric acid methyl ester (PCBM).[7-9] In a typical BHJ OPV architecture, this photoactive layer is sandwiched between two electrodes and eventually additional charge-selective interlayers in-between to enhance the collection of the photogenerated charges and decrease leakage currents.[10-12] Today, record efficiencies of lab-scale OPV devices are reported in the range of 10-12%.[13] However, irreproducibility and instability are still major problems of this type of devices, which lead to performance variations up to 10% batch-to-batch on identical devices. Therein the operation environment has been identified to play a critical role as potential source for OPV degradation, with four main factors primarily responsible for limiting the stability: oxygen, water, light and heat.[14,15] Deterioration by photooxidation of the conjugated polymer, diffusion of foreign atoms creating charge traps and recombination centers, corrosion of the electrodes or thermally induced morphological reorganization are only a few examples of the manifold effects occurring.[16,17] In some cases, degradation is even accelerated by introducing certain auxiliary layers into the device architecture. Prominent example is the prominent hole-transport material poly(3,4-ethylenedioxythiophene):poly(styrenesulfonate) (PEDOT:PSS), once popular for its advantageous solution processability, high transparency, work function and conductivity.[18,19]

When applied between the inorganic indium-tin-oxide (ITO) electrode and the organic semiconductor, its function is to block electron transport and lower the hole-extraction barrier to the anode. Unfortunately, the hygroscopic acid component PSS causes electrode corrosion in humidity, leading to indium diffusion, instability, shorter life times and spatial inhomogeneity of devices.[20-23] To circumvent these problems, alternative interlayers such as transition metal oxides[24-26] have been used to substitute PEDOT:PSS as HT layers in OPVs. Despite the fact that numerous long-term degradation and stability studies with standard OPV device architectures have been reported, only few publications are dedicated to degradation and homogeneity issues free of the influence of HT layers.[27]

This work aims to fill this gap by studying short-term environmental effects on standard P3HT:PCBM OPV devices in absence of any HT layers between ITO electrode and photoactive layer. The architecture of the devices studied is shown in Fig. 1. Thereby a comparison was made between exposure to different environments: in argon, dry air or humid air in the dark, and in dry air under illumination. Different sorts of imaging techniques have been used in the past to monitor degradation of optoelectronic devices.[28-30] In photovoltaics, laser beam induced current (LBIC), is a prominent method for mapping the photocurrent response across the device area and thus localizing fabrication/degradation defects.[12,31-34] Additionally, local photocurrent-voltage characteristics at different sites of the device can be used to identify the nature of the defect causing the locally diminished photocurrent. This technique is adapted from photovoltaic module testing, where the physics of the entire module is determined by the physics of assembled subdevices.[35] In the case of a single solar cell, the spatial distribution of heterogeneously responding areas can be seen as parallel-connected microdiodes,[36] which has been demonstrated for polycrystalline inorganic thin-film solar cells e.g. by Karpov et al.[37]

By connecting global device behavior with such spatially resolved characteristics and photocurrent imaging, we gain knowledge about the predominant factors causing initial performance losses of the devices shortly after fabrication.

EXPERIMENTAL SECTION

**Materials**

Poly(3-hexylthiophene) was supplied by Rieke Metals Inc. (MW 50-70 kg mol$^{-1}$, regioregularity 91–94%). [6,6]-phenyl-C$_{61}$-butyric acid methyl ester was purchased from Nano-C Inc. (99.5% purity). Anhydrous chlorobenzene, as solvent for the organic semiconductors, was purchased from Sigma-Aldrich. All materials were used as received. Indium-tin-oxide coated glass substrates (20 Ω /square, Ossila Ltd.) were cleaned by sonication in acetone and isopropanol, followed by O$_2$-plasma etching (100 W for 30 min) briefly before use.

**Sample preparation**

Standard P3HT:PCBM solar cells without HT layer were fabricated with the layer structure ITO/P3HT:PCBM/LiF/Al. The active layer was applied directly onto patterned ITO glass by spin-coating from a solution of P3HT and PCBM (1:1 weight ratio, each 18 mg/mL in 70°C chlorobenzene) at 1500 rpm for 60 s, followed by annealing at 120° C for 5 min. The obtained active layer thickness was around 150 nm. The cathode was thermally evaporated as a bilayer of LiF (2nm)/ Al (100nm). Active layer samples for photophysical testing were prepared as described above on spectrosil quartz substrates. For mimicking the different environmental conditions, the non-encapsulated devices were exposed to static argon (Ar), dry air with < 20% relative humidity (Air) or humid air with > 80% relative humidity (H$_2$O) without any illumination and to dry air under white light illumination (tungsten halogen, 10% equivalent of AM1.5G) (Air & light), respectively. Durations of exposure were varied between 0 and 240 min and devices characterized immediately after. Active layer samples were treated accordingly. The ambient temperature was around 25 °C for all conditions. To avoid additional effects from long-term photocurrent mapping in air, exposed devices were encapsulated before measurement.

**Characterization**

UV-Vis optical absorbance spectra of the P3HT:PCBM active layers have been recorded between 300 and 800 nm, using a Shimadzu UV-1800 UV–Vis spectrophotometer. Fluorescence spectra were acquired on a Fluorolog3 spectrofluorometer (Horiba Jobin Yvon) equipped with a NIR-sensitive photomultiplier R2658 from Hamamatsu (300–1050 nm). Excitation wavelength was 530 nm. The absolute luminescence quantum yields were determined using an integrating sphere from Horiba. Both UV-Vis absorbance spectra and fluorescence spectra were recorded in air. The photocurrent density–voltage (J–V) characteristics of the solar cells without encapsulation were measured in argon (glovebox) under simulated solar illumination (AM1.5G) at 1 sun (100 mW/cm$^2$) (solar simulator, ABET Technologies, Model 10500) using a computer-controlled Keithley 2636A source meter. Their spectral response was recorded from external quantum efficiency (EQE) acquired for wavelengths from 375 to 900 nm, using a 250 W white light source (tungsten halogen) with monochromator, with a computer-controlled Keithley 2636A source meter and calculated against a calibrated silicon photodiode. Spatial photocurrent distribution was determined by laser beam induced current (LBIC) method, scanning with a computer-controlled nano-manipulator-driven xy-stage (Kleindiek Nanotechnik, NanoControl NC-2-3) and excitation from the glass/ITO side with a 532 nm laser (< 5 mW) with a focused spot-size of ≈2 μm. For spatial photocurrent mapping, local short circuit current was scanned with 40 μm step size across the entire photoactive pixel area of 4.0 mm x 1.5 mm (Figure. 1). Local photocurrent-voltage characteristics for point-to-point analysis were taken in line-scans along the long side of each pixel, scanning from the left (0 mm) to the right (4 mm) with 200 μm step size. EQE, LBIC and local photocurrent-voltage characteristics for point-to-point analysis were performed on encapsulated devices to prevent avertible further effects by measurement in air atmosphere.

## RESULTS AND DISCUSSION

**Integral device behavior**

To understand the type and dynamics of how different environments affect OPV devices, the repeatedly exposed devices where analyzed depending on duration and atmosphere. The additional short time light exposure during measurement was confirmed to be negligible for demonstration of the relevant effects in this study. The solar photocurrent density-voltage (J-V) characteristics of the devices after exposure sequences of 60 minutes up to a total duration of 240 minutes are shown in Figure 2a. The respective solar cell key values, such as short circuit current density ($J_{SC}$), open circuit voltage ($V_{OC}$), fill factor (FF), power conversion efficiency (PCE) and series resistance ($R_S$) were extracted/calculated from the according characteristics and summarized in a plot in Figure 2b for better visibility of even small changes. Before exposure, all devices show identical characteristics (Fig. 2a) and key values (Fig. 2b), with $J_{SC}$ of almost 10 mA/cm$^2$, $V_{OC}$ of 0.5 V, FF of almost 60% and a resulting PCE of about 3%, confirming their comparability. The argon test series acts as the blind sample in this study, to account for environment-independent material relaxation and aging, but also for effects by measurement conditions. This is necessary, because for P3HT:PCBM solar cells in particular, significant morphological changes to the active layer have been observed during operation,[38] different to the same active layer within an inverted configuration.[39] Accordingly, the J-V characteristics of the devices that were exposed to argon (Fig. 2a) show expectedly least variation over exposure time, with equal shape of the curves, slopes and $V_{OC}$. Merely an insignificant decrease in photocurrent $J_{SC}$ is noticeable, whose cause is suspected in the minor decay of the active layer due to repeated applied voltage under illumination during characterization. The lowered $J_{SC}$ in consequence leads to slightly lower efficiency PCE, while all other values are stable (Fig. 2b, black squares).

Compared to that, not illuminated air-exposed solar cells (Fig. 2a) show slightly stronger reduction of photocurrent over exposure time. Looking at their key values (Fig. 2b, red circles), it can be seen that here the $V_{OC}$ is weakly affected and the series resistance is slightly increased. This indicates a slightly accelerated aging process, compared to samples stored in inert gas. If devices are exposed to air under illumination, a condition commonly feared to cause detrimental photooxidation of the active layer, here merely the same scale small variations as for the air/dark sample are seen (Fig. 2a). In comparison with the air/dark series, the key values of the illuminated air devices (Fig. 2b blue triangles) show identical development of the $J_{SC}$ over time, also only minor changes to $V_{OC}$, but differ presenting a small decrease in fill factor and an increase in series resistance. However, this still very similar behavior indicates that low intensity light sources (here 10% intensity of 1 sun) do not lead to further device degradation. In contrast to that, devices that were exposed to humid air without illumination show detrimental development with exposure duration. This is immediately visible from their J-V curves (Fig. 2a) showing considerable reduction of photocurrent with increasing exposure time. Thereby $J_{SC}$ dramatically decreases by 25% within the first 60 minutes and continues to decay by up to 75% after 240 minutes exposure. By looking at the key values of the devices of the humidity series (Fig. 2b, green stars), an observation is that the significant increase in series resistance with exposure time show strongest direct effect on $J_{SC}$, while the related FF and the $V_{OC}$ are relatively little affected. With the $J_{SC}$ as main humidity responsive factor contributing to PCE, also the efficiency follows the trend of photocurrent development. The relatively small response of $V_{OC}$ and FF to humidity implies that the effective bandgap of active layer stays unchanged and no major leakage currents were generated after exposure. The trend of all factors ($V_{OC}$, $J_{SC}$, FF, $R_S$, and PCE) for the devices with Ar/dark, air/dark and air/light exposure are relatively similar. The slight efficiency losses that occur are dominated by drop of $J_{SC}$, which corresponds to the rise of series resistance ($R_S$). In the case of the device series with humidity exposure, $R_S$ increases considerably, giving rise to the suspicion that humidity led either to degradation of the bulk active layer material or generation of interfacial barriers at the electrodes. To clarify which major mechanisms are responsible, further characterization was acquired and discussed in this paper. Numerous past publications refer to oxygen/humidity-based degradation of the organic/electrode interface in OPV devices, with quite differing arguments about the actual role of water alone. Often observed are voltage-dependent interfacial charge-transfer barriers, typically recognizable from S-shaped J-V curves with lowered fill factor and reduced $V_{OC}$.[15,16,40-43] However, in the present case, J-V curves are not deteriorated and both, FF and $V_{OC}$ are barely affected, indicating that the interfacial charge-transfer efficiency between the active layer and electrode is unaltered during exposure. Further, the J-V curves of all exposure conditions show still parallel slopes at the intersection with the current axis ($J_{SC}$ region) with increasing aging time, implying that shunt resistance ($R_{SH}$) stays also unchanged, therefore treatment had also no effect on the magnitude of leakage currents of the device.

**Photophysics and photoresponse**

Beside the magnitude of $J_{SC}$, also the spectral response of a solar cell can carry valuable information on their physical properties. The external quantum efficiency (EQE) is determined by $J_{SC}$ per wavelength and is a measure for the spectral response of the device. While the total magnitude of EQE follows the trend in photocurrent as presented before, the normalized EQE spectra allow further assumptions by visualizing potential spectral deteriorations. Figure 3 shows the normalized EQE spectra of devices following 240-minute exposure at different conditions.

The spectra of those solar cells exposed to Ar/dark, air/dark and to air under illumination, all exhibit the congruent signature shape for EQE of P3HT:PCBM OPV devices, which is a broad peak from 380 nm to 700 nm with a maximum at 500 nm and a subtle shoulder at 620 nm, basically mirroring P3HT absorption (compare also Fig. 4a). Only upon humidity/dark exposure, a deviation of the EQE spectrum occurs, noticeable as a flattening of the peak at the long wavelength side. Potential reasons for this effect would be degradation either of the active layer material or the organic-electrode interface. In the first case, degradation would have altered the absorption spectrum of the active layer material. In the second and (as becomes clear in the following) more likely case, the flattening at long wavelength is caused by diminishing number of charge carriers created in the back part of the solar cell, close to the LiF/Al cathode. The reason for this is that long-wavelength radiation has larger penetration depth in P3HT:PCBM, therefore a large contribution of charges is coming from the entire volume of the active layer, while at the 500 nm peak absorption, penetration depth is small and thus contributions are mainly from the front of the cell, close to the ITO anode. Similar effects have been reported in the past, most prominent the M-shape deterioration of EQE, caused e.g. by self-absorption by too thick active layer, or in cases of degradation or barrier behavior near the transparent anode.[44] In the present case, the loss of charge carriers at higher wavelengths therefore could indicate degradation of active layer near the electrode or the organic/cathode interface.

Monitoring the absorption and emission behavior of conjugated polymers has proven to be useful to recognize changes e.g. due to photobleaching effects caused by photooxidative degradation.[45-49] To investigate how exposure environments affect the active layer alone, its UV-Vis absorption and photoluminescence (PL) properties after 240 min exposure to the respective atmospheres have been investigated and compared to the pristine state. According spectra are shown in Figure 4. The absorbance of the active layers (Fig. 4a) shows the expected features of both blend components, which is the absorption of PCBM in the UV region below 350 nm and the characteristic P3HT peak at 500 nm with the small vibrational shoulder at 620 nm. In comparison, the spectra look identical, beside a negligibly higher general absorption of the pristine sample. The emission behavior of a conjugated organic molecule is an even more sensitive marker to spot degradation. However, looking at shape of the normalized PL spectra (Fig. 4b), again all samples exhibit the typical P3HT features in form of the 0-0 emission at 650 nm and the 0-1 emission at 720 nm with no significant shift of the peaks or changes of the I0-0/I0-1 ratio. In addition, the absolute photoluminescence quantum efficiency (PLQE) of the films was determined with 0.33% for fresh samples and 0.29% for all exposed samples with no difference between argon, air, illumination or humidity treatment. This rather small change between fresh and any aged sample, with no relation to the applied different environmental conditions, origins most likely from relaxation of the film. Therewith the photophysical results confirm that the applied exposure processes in this study do not harm the active layer, at least if separated from the rest of the stack, i.e. without contact to electrode material.

**Spatially resolved device characteristics**

To gain an understanding about the mechanism behind the loss of photocurrent, which seems to occur without apparent active layer degradation or interface barrier formation, the spatial homogeneity of the device physics across the solar cell area was investigated. Therefore, overview maps of the local photocurrent of respective solar cells after different environmental exposure were generated by LBIC scanning at 532 nm illumination. The normalized photocurrent density maps are shown in Figure 5 for 60 min (a) and 240 min exposure (b), respectively. For reference, an according map of a fresh device before exposure is provided in the Supporting Information in Figure S1. The device stored in inert Ar atmosphere in the dark shows expectedly a very homogeneous high output area, independent of the exposure time. Merely at the edges of the pixel area and a few central spots (≈ 90 to 130 μm in size), the photocurrent is lowered to around 90 % of the original performance, which might be due to regular morphological inhomogeneities of the devices.[12] These images represent the basic spatial performance without further environmental influence. Upon exposure to air/dark and air under white light illumination, a very similar pattern is found: Compared to argon-stored devices, the number of low-current-spots increases by a factor of two for both conditions, while average lateral dimension of the spots remains constant. The extent of deficiency of most spots is similar to the ones found in argon samples, still reaching around 90% of the original performance, but there are also few spots where photocurrent is reduced down to 80% after 60 min or even below after 240 min. This means that there is a small effect caused by oxygen exposure but no significant effect seen from additional illumination. A completely different picture is drawn by exposing the devices to humidity in the dark. After only 60 min exposure, the solar cell area shows spots and larger islands of failing photocurrent output, almost down to less than 10 % of the original performance, while the other parts of the cell show unchanged high output of 100%. The density of low-output zones has been calculated to be 1/5 of the pixel area. At exposure beyond 60 min, this humidity caused local device failure continues to propagate. After 240 min, there are only few islands of unaltered high photocurrent output left, while the majority of the area is characterized by a quite homogeneous low-output zone with less than 10 % of the original photocurrent. Here, a calculation of the low-output area makes about 3/5 of the device.

The observation of island-like device failure upon humidity exposure is different to previous reports e.g. by Madogni et. al,[43] showing only for oxygen the formation of islands by growth of $Al_2O_3$ at the metal/organic interface, but instead homogeneous (photo)degradation caused by $H_2O$. However, since many studies in the past were commonly made on devices containing also the hygroscopic hole-conductor PEDOT:PSS, the question is if on those occasions PSS-caused corrosion might have overshadowed other humidity effects.

Probing and comparing the local device physics in areas of high and low output, might allow isolating the cause for locally lowered photocurrent and thus help to identify potential bulk material or interface degradation mechanisms connected to it. For that purpose, local photocurrent-voltage characteristics were recorded across the area of 240 min-exposed solar cells for point-to-point analysis. Figure 6a shows the local photo-J-V curves from a central line-scan along the long-side of 4 mm solar cell pixels for the four conditions Ar/dark, air/dark, air/light and humidity/dark. The solar cell parameters $J_{SC}$, $V_{OC}$, FF, $R_S$ and PCE, as extracted/calculated from those photo-J-V characteristics, are plotted over their position on the pixel in Figure 6b. For better visibility of any performance scattering between local J-V characteristics across a device area (Fig. 6a), the curves have been color-coded in regard of their achieved photocurrent. Thereby the "devices' best curve", is represented by a dark red line, with a $J_{SC}$ of around -0.22 mA/cm$^2$ for all conditions and then following the rainbow in sensible steps of decreasing $J_{SC}$ (bright red, orange, yellow, green, turquoise, light blue, dark blue, purple). As expected from the photocurrent maps (Fig. 5), the device stored in argon without illumination exhibits the the smallest deviation from the "best curve" and also no gradual performance scattering, but a majority of points on the device with maximum performance and only few singularities with slightly lowered photocurrent, but only down to -0.21 mA/cm$^2$ for the worst point. Again, air/dark and air/light exposure lead to very comparable results, with a larger deviation from the "best curve", compared to argon/dark. Still, it is visible that the majority of the measured points deliver maximum performance and only some singularities, but more than for Ar/dark, show lowered photocurrent, down to a $J_{SC}$ of -0.17 mA/cm$^2$. This agrees well with the impression from the photocurrent map (Fig. 5). Also for the latter three conditions, the $V_{OC}$ remains unchanged and, slopes are stable, which makes any changes to charge injection/extraction or leakage, unlikely as reasons for the diminished photocurrent. Contrary to the previous conditions, exposure to humidity without illumination leads to tremendous scattering of the local J-V-curves with $J_{SC}$ varying between -0.22 mA/cm$^2$ and -0.01 mA/cm$^2$. Thereby also the population of curves with best performance is considerably decreased, while a larger number accumulates at a $J_{SC}$ of -0.06 mA/cm$^2$ instead. The trend between the four conditions becomes even more obvious, when comparing the normalized (to each solar cells' maximum) location-resolved plotted solar cell parameters (Fig. 6b). Please note that also the resistance is normalized to the devices' overall maximum, meaning a constant "high" $R_S$ does not indicate a high resistance but merely a very small magnitude of change across the device area. As expected from photocurrent mapping (Fig. 5), the Ar/dark device exhibits clearly the highest spatial homogeneity, with very constant values for $J_{SC}$, $V_{OC}$, FF, $R_S$ and PCE across the entire area, except for a slight drop of $J_{SC}$ (and with it in PCE) at the pixel edges.

The latter is caused by the non-sharp border of the pixel area, which is defined by the dimensions of the evaporated LiF/Al electrode patch. Vapor deposition through a shadow mask is usually accompanied by formation of a small diffuse zone (gradually fading thickness) at the edges, which here might lead to gradually lower photocurrent depending on the location at or the distance from the cathode patch.[50] Consistent with the photocurrent maps and local J-V-curves, the devices exposed to air/dark and air/light show also in the parameter plot comparable patterns. Thereby $J_{SC}$, $V_{OC}$, FF, $R_S$ and PCE are almost constant across the area, except for some singularities, which show a lowered photocurrent (and PCE) but other values remaining constant. Both devices also show the edge effect observed in the argon/dark sample, even slightly more pronounced by a slightly stronger decreased photocurrent. In contrast to those three conditions, the humidity/dark exposed sample shows a dramatic pattern in the parameter plot, with strong local variations. Thereby regions of barely affected high $J_{SC}$, $V_{OC}$, FF, PCE and relative low $R_S$, are neighboring areas with heavily increased resistance $R_S$, $J_{SC}$ almost lowered to zero and further not as strongly affected, but still lowered $V_{OC}$ and FF. Also for humidity exposure, edge effects are visible and seem more heavily pronounced than for the other samples. Though one might assume that this originates from humidity diffusing underneath the "impenetrable" LiF/Al patch, the photocurrent maps (Fig. 5) of the humidity/dark device, exhibiting also barely affected regions along the edges, rather suggest this being due to an overlap of the edge with a "degraded region" within the scan range. These findings correlate very well with the observed integral device performance for the different environmental conditions (Fig. 2). Despite the fact that for the integral performance an AM1.5G standard white light source was used and a green laser diode for the photocurrent scans, the trend of the overall device performance is obviously a consequence of the summarized local properties. Question is still which chemical/physical processes lead to the observed local behavior. Here, literature gives us some known facts and options, which parameters of a solar cell are usually affected by which kind of mechanisms[40,43]: A lower FF indicates reduced interfacial charge transfer efficiency between different layers, e.g. active layer and electrode interface. Higher series resistance is usually caused by either larger contact resistance at the active layer/electrode interface or by deterioration of the bulk active layer material. A decrease of $V_{OC}$ may be a sign for a reduction of active layer/metal electrode interface, a lowered effective band gap of the donor/acceptor couple (e.g. by material degradation) or a change of electrode work function. The drop in $J_{SC}$ may stand for degradation of the active layer, which decreases the photon absorption, alter the charge mobility and exciton dissociation efficiency, or/and deterioration of the electrode/active layer interface, limiting charge carrier transport and collection efficiency.[40] A definite answer if the observed behavior comes from degraded active layer at contact with an electrode (because pure active layer degradation was already excluded earlier from photophysical characterization) or from an interface degradation, the charge transport characteristics of the exposed ITO/P3HT:PCBM/LiF/Al devices were studied.

**Charge transport**

Charge mobility of devices has been extracted by modeling the dark current under forward bias using the space-charge-limited expression for the current density[51]:

$$J_{SCL} = \frac{9}{8}\varepsilon\mu_0 e^{0.891\gamma\sqrt{V_{int}/L}} \frac{V_{int}^2}{L^3}$$

where $\mu_0$ is the zero-field mobility, $\gamma$ the field activation parameter, $J_{SCL}$ the current density, $\varepsilon$ the permittivity, $V_{int}$ the internal voltage, and L the film thickness. Assuming that the barrier for injection of electrons into the PCBM LUMO from the Al electrode is higher compared to the injection of holes into the P3HT HOMO from ITO, these devices will be hole-dominated under forward bias, therewith the values are expected to be close to the hole mobilities in the P3HT:PCBM blend. While the absolute values might be higher than in a true hole-only device,[52] the values will allow a relative comparison between the different exposed devices. Space-charge limited current characteristics and the respective fits of devices after exposure to Ar/dark, air/dark, air/light and humidity/dark are shown exemplarily for 240 min exposure in Fig. 7. The complete table of fitting parameters and derived charge mobilities at exposure of 60 minutes and 240 minutes, respectively, is given in the supplemental. Beside the devices' constant parameters L and ε, also $V_{bi}$ and fitting parameter γ were quite comparable between different devices and conditions. First, the mobilities were fitted for all devices using the current density for the nominal device area of 6 mm$^2$. For 60 min exposure, the zero-field mobilities were 1.29 x 10$^{-7}$ m$^2$/(Vs) for Ar/dark, 1.24 x 10$^{-7}$ m$^2$/(Vs) for air/dark, 1.31 x 10$^{-7}$ m$^2$/(Vs) for air/light and 9.5 x 10$^{-8}$ m$^2$/(Vs) for humidity/dark. While the values for argon and air, with or without illumination, are equal within the level of accuracy and comparable with common values for P3HT:PCBM devices,[53] the humidity-exposed sample shows a by 25% reduced charge mobility. However, following the observations from photocurrent mapping that a fraction of 1/5 of the 60 min humidity/dark-exposed device area was failing to contribute, question is if the material or the contact area is affected. Therefore we repeated the fit with an according area-corrected (80% of 6 mm$^2$) current density and obtained a mobility of 1.20 x 10$^{-7}$ m$^2$/(Vs), therewith within the same range of the other devices. After an exposure time of 240 min at the different conditions, the obtained mobilities were 1.46 x 10$^{-7}$ m$^2$/(Vs) for Ar/dark, 1.24 x 10$^{-7}$ m$^2$/(Vs) for air/dark and 1.31 x 10$^{-7}$ m$^2$/(Vs) for air/light, which means that for these three conditions the longer exposure does not have a significant effect on the mobility. Only for humidity/dark, the mobility is even lower than before with only 5.96 x 10$^{-8}$ m$^2$/(Vs) for the nominal device area. However, the fit by using only the intact fraction of the device area, which was estimated from photocurrent mapping with 2/5 (40% of 6 mm$^2$), the obtained mobility is with 1.19 x 10$^{-7}$ m$^2$/(Vs) again comparable to other exposure conditions. These results support the image of a stable and unchangeable performance of active layer independent of environment and duration. Therefore the aforementioned lowered $J_{SC}$, $V_{OC}$ and increased $R_S$ is not induced by active layer degradation in the present case.

**Discussion**

The presented results show clearly that exposure to dry air, whether illuminated or not, does only induce minor efficiency decrease of P3HT:PCBM devices without hole transport layer, compared to those stored in argon in the dark. This subtle decrease is caused by singularities on the device area of slightly diminished photocurrent, while $V_{OC}$, FF and $R_S$ are not affected in these positions. As these point defects have about the same lateral dimensions as seen for Ar/dark devices, even if these show negligible current fluctuations in these spots, it is suspected that these are material defect sides present in any device, e.g. foreign atoms or particles, which might act as potential reaction sites when getting into contact with oxygen. This is supported by the fact that the size, density and intensity of these spots is independent of exposure time or illumination. Accordingly, also no barrier behavior, active layer photodegradation or mobility decrease is noticeable in these devices. Exposure to humidity in the dark on the other side, obviously triggers serious efficiency losses in the HTL-free solar cell. From pure photophysical point of view, the P3HT:PCBM blend showed to be stable under the applied conditions, even after long exposure times, which suspends any significant water-related degradation of the active layer itself. However, this does not eliminate the possibility of interface reactions of the organic donor/acceptor catalyzed by or with the electrode material. On this occasion EQE spectra, which join photophysics with device physics, indicated a diminished contribution of charge carriers at the LiF/Al side of the humidity-exposed device, which identifies the "problem side" of the device. This is clearly different to PEDOT:PSS containing standard solar cells, where humidity-caused degradation is majorly induced at the anode side and often ascribed to ITO corrosion by PSS and/or its hygroscopic behavior, fostering water uptake and in the cause of layer swelling, delamination. The present disturbed photocurrent generation on the aluminum side could have its origin in limited charge transfer from the organic to the LiF/Al cathode or in diminished generation/transport of charge carriers within the active layer close it. In literature, water-induced degradation was associated with the formation of interfacial oxide and S-shaped current-voltage characteristics. Though we can see island-like failure areas of diminished photocurrent and they do show subtle drop in $V_{OC}$ and FF in local in the local scan, we do not observe any classical indications for voltage-dependent barrier formation near the electrode, nor failing charge transfer, which would result in a deteriorated current-voltage characteristic. Instead, a risen series resistance in the local failure regions seems to weigh much stronger. This is also supported by the charge mobility evaluation, which clearly shows intact unchanged transport behavior, if the "failure areas" are neglected. Therefore it is assumed that the active layer, even in contact with the cathode, is still widely intact. But the charge carriers generated in such a region have to be transported a longer distance to an intact electrode area to be extracted, which leads to a virtually higher resistance and the noticeable but small local FF and $V_{OC}$ decrease. In consequence, the properties of partial contributions from regions with disturbed contact area, do not have significant impact on the total integrated device properties, except for the loss of photocurrent. Therefore no change in FF and $V_{OC}$ is visible for the integral device performance of the humidity/dark device. However, in literature the opinions about the role of humidity or oxygen in organic solar cells are contradictory or no differentiation is made at all. A report by Hermenau et al.[54] is one of the few, which documented them separately. Thereby they found that both, molecular oxygen and water, diffuse through the aluminum electrode. According to their results, oxygen diffused rather through selected pinholes in the aluminum film, forming $Al_2O_3$ islands near the organic interface as consequence. For water, they suggested diffusion along the grain boundaries of the evaporated aluminum layer, resulting in rather homogeneous degradation of the device. While we can definitely confirm their further observation that humid air seems more reactive than dry molecular oxygen, which is well known from metal corrosion, we cannot confirm the homogeneous degradation by water. Therefore, we suggest that in the present case water diffusion is also rather subject to pin hole transport. We assume that at the bottom of this pin hole, it might react to form an oxide or hydroxide, but apparently to such an extend (regarding its thickness) that it cannot be overcome by an applied voltage and rather lead to completely insulated areas. The presence of LiF underneath the aluminum might also cause some effect. Though it is still not entirely clear why a sub monolayer of LiF between the organic acceptor and aluminum electrode improves the devices, it is a known fact that LiF added to aluminum melts acts as oxidation inhibitor by the formation of AlF at the surface. It is very likely that also in this case this could have prevented the formation of interfacial oxide in air/dark and air/light devices in this study. In humidity, the LiF layer might get partially dissolved, exposing a fresh aluminum surface which then would quickly react with water via $Al_3+ + H_2O \longrightarrow Al(OH)_3 + 3H^+$. This might not only cause formation of an oxide/hydroxide island but also, due to the production of hydrogen, lead to delamination of the electrode by formation of bubbles.

CONCLUSIONS

We aimed to gain information about the occurring degradation mechanisms, in organic solar cells without involvement of acidic PEDOT:PSS, which in standard solar cells dominates the degradation discussion. Therefore we investigated the impact of chosen environmental exposure conditions on the performance of devices with the architecture ITO/P3HT:PCBM/LiF/Al, i.e. without any hole-transport layer between ITO and organic active layer. Among the different conditions, argon storage acted as the blind sample and was compared to exposure to dry air in the dark or with illumination, and to humid air in the dark, with different exposure durations up to 240 min. We found that the photophysics of the active layer stays unchanged for all applied conditions and giving no signs for photodegradation. In general, the exposure to oxygen (air) did not lead to significant performance losses. The small difference to argon-stored samples seems to originate from small device defects, which are also visible in traces in argon-stored devices.

Devices exposed to humid air on the other hand did show huge performance losses during the same exposure times. EQE spectra identify the aluminum side as the primary suspect and LBIC scans show time-dependent formation of patches of significantly reduced photocurrent contribution, but instead of usual interfacial charge extraction barriers, rather insulating behavior is found. Different previous reports, it is suggested that not only oxygen, but also water is transported through pin holes in aluminum to the organic/electrode interface, leading to the formation of insulating islands and not homogeneous device degradation. The fact that no island formation is seen for dry air exposure is assigned to the protective properties of the LiF interlayer.

## ASSOCIATED CONTENT

**Supporting Information**

The Supporting Information contains a table summarizing constants, fitting parameters and charge mobilities as derived from space-charge limited current modelling on photovoltaic devices exposed to Ar/dark, air/dark, air/light or humidity/dark for 60 min or 240 min, respectively. Additionally, an LBIC map of a fresh device before exposure is provided for reference.

AUTHOR INFORMATION
**Corresponding Author**
*Email: hchien@tugraz.at
**Notes**
The authors declare no competing financial interest.

## ACKNOWLEDGMENT

H.-T. C. and B.F. are grateful to Sergey M. Borisov for the access to equipment and valuable discussions, and to the Austrian Science Fund (FWF) for financial support (Project No. P 26066).

FIGURES

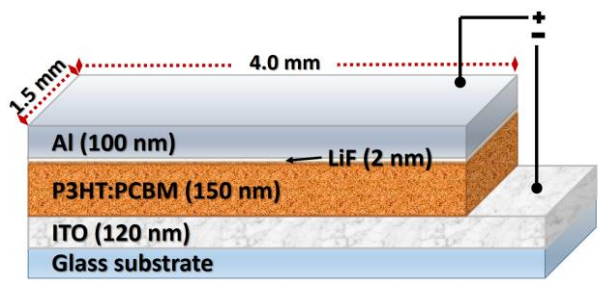

**Figure 1.** Illustration of the device architecture for hole-transport layer free organic solar cells investigated in this study.

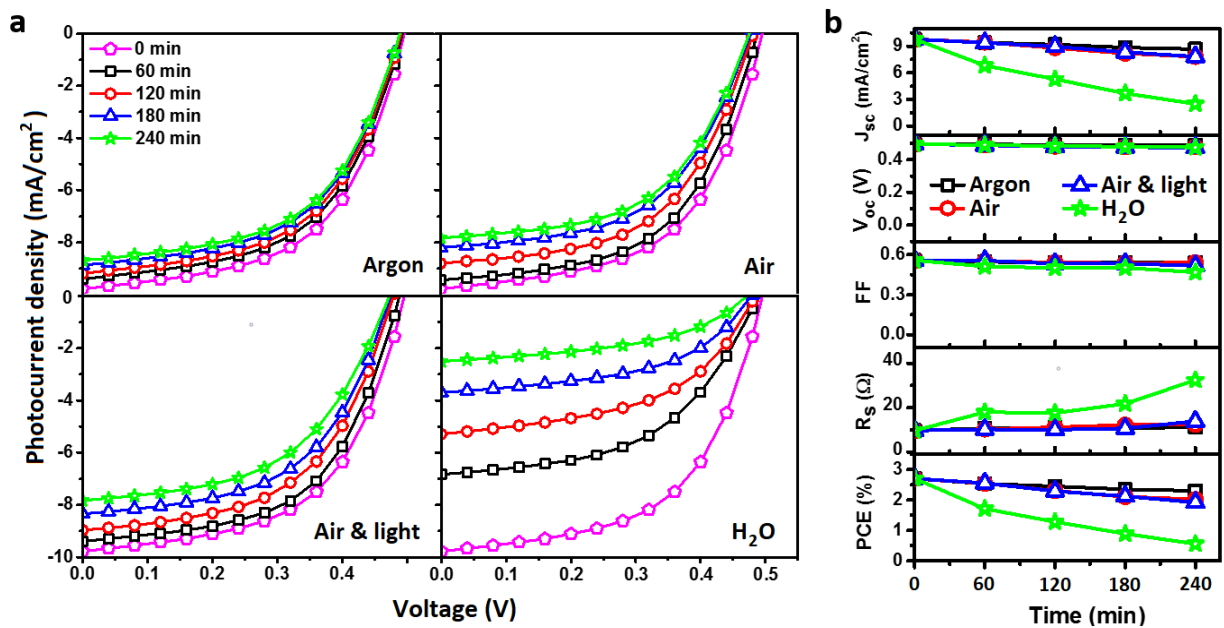

**Figure 2.** (a) J-V characteristics of devices changing from pristine state over repeated 60 min environmental exposure sequences up to 240 min in argon/dark, air/dark, air/light, and humidity/dark exposure. (b) Overview of solar cell key values changing with increasing exposure time at different environmental conditions.

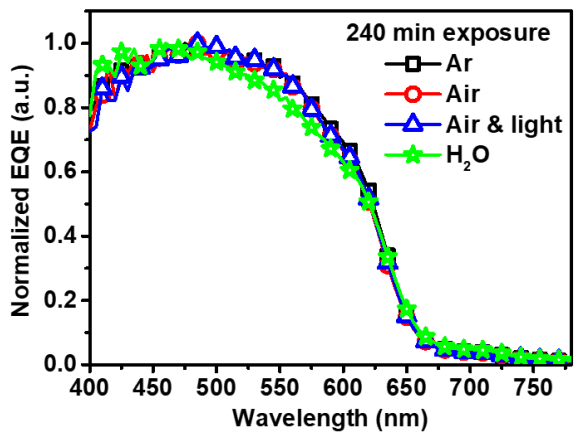

**Figure 3.** Normalized EQE spectra of devices following 240 min exposure in different environments.

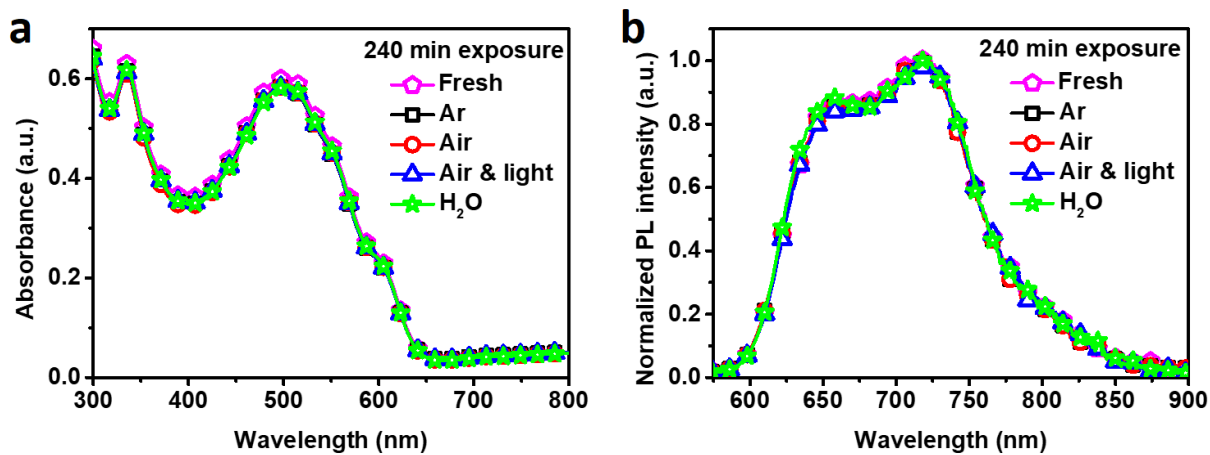

**Figure 4.** UV-Vis absorbance (a) and normalized emission spectra (b) of P3HT:PCBM films pristine or after 240 min exposure to different environments.

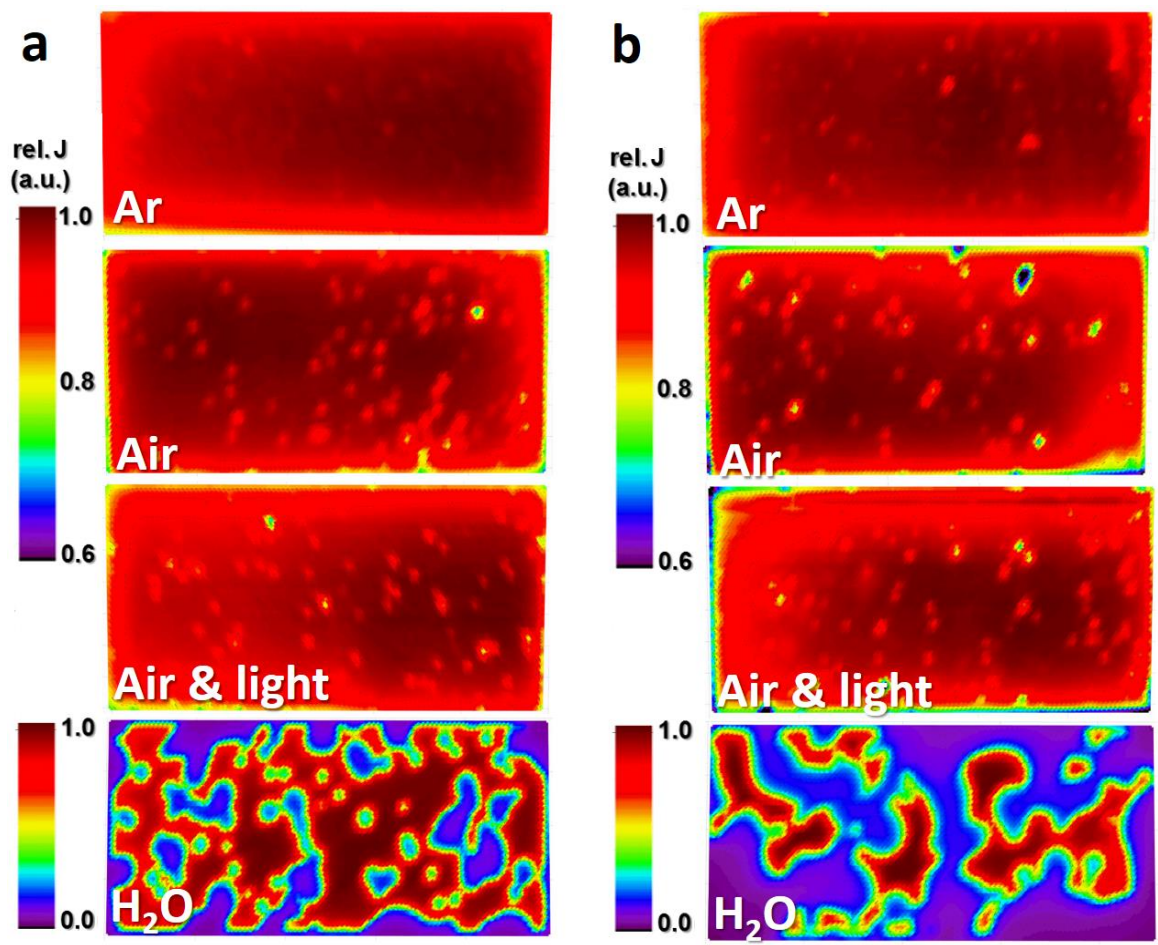

**Figure 5.** LBIC maps of device area (1.5 mm x 4.0 mm) upon exposure to different atmospheres for 60 min (a) and 240 min (b).

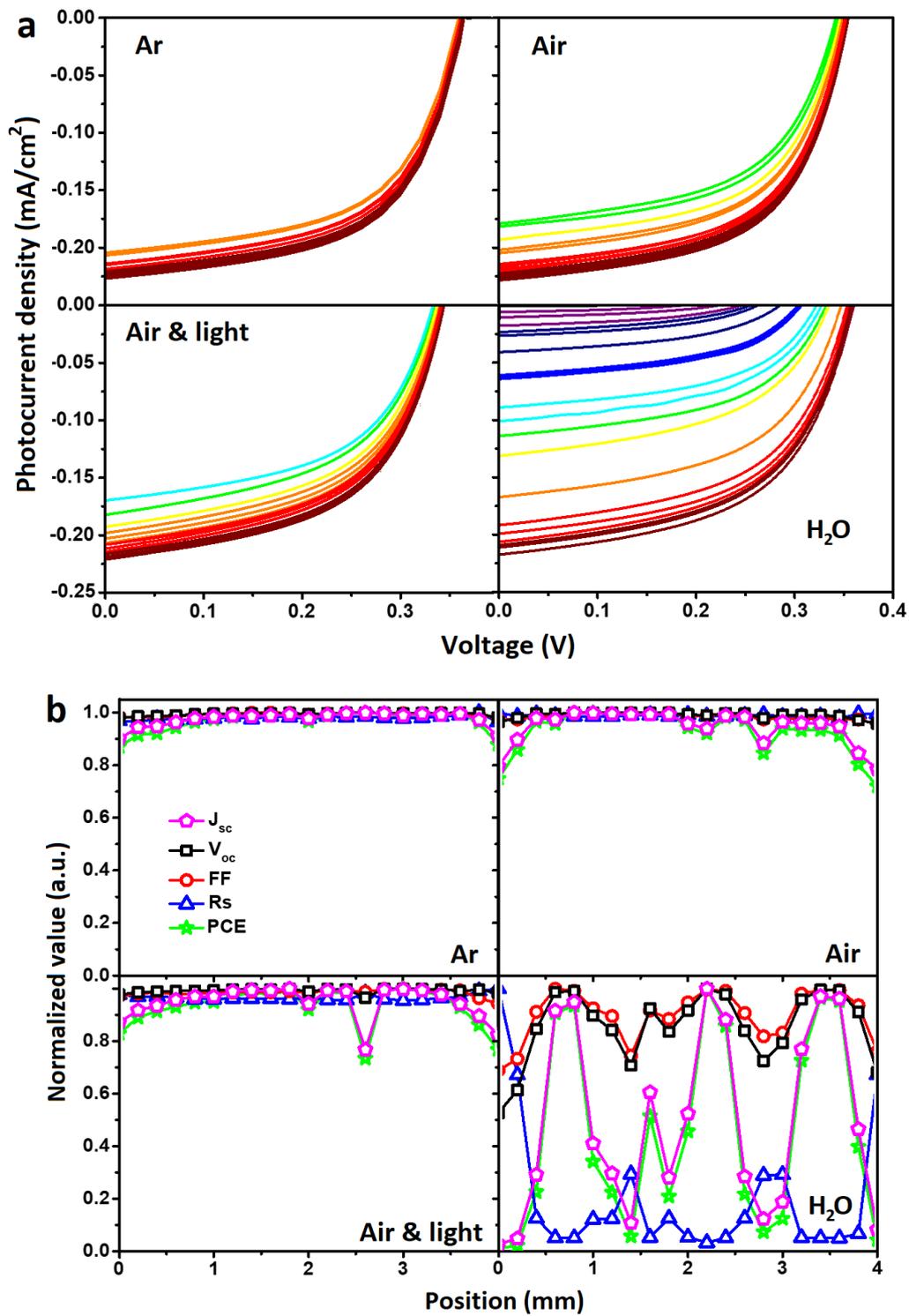

**Figure 6.** Local photo-J-V curves recorded at different points along the long-side (4 mm) of the 240 min-exposed (Ar/dark, air/dark, air/light, humidity/dark) solar cells (a) and the respective parameters extracted from the local photo-J-V characteristics plotted over the scan-position on the pixel (b).

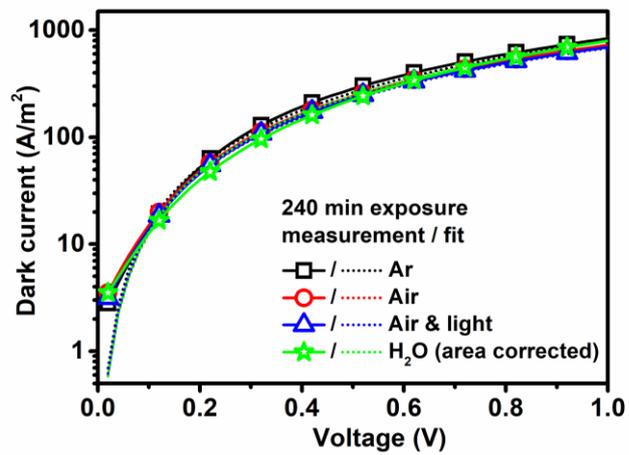

**Figure 7.** Exemplary current-voltage curves of 240 min-exposed ITO/P3HT:PCBM/LiF/Al devices (solid lines with symbol) with according fits to the space-charge limited model (dotted lines).

SYNOPSIS TOC

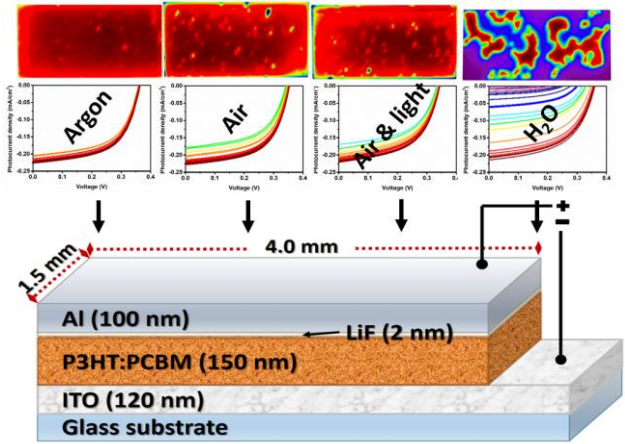